\documentclass[aps,nofootinbib,preprintnumbers,tightenlines]{revtex4}
\usepackage{epsfig}

\def\bea{\begin{eqnarray}}
\def\eea{\end{eqnarray}}
\def\bec{\begin{center}}
\def\ec{\end{center}}

\def\beq{\begin{equation}}
\def\eeq{\end{equation}}

\def\p{\partial}
\def\f{\frac}

\def\f#1#2{\frac{#1}{#2}}
\def\p{\partial}

\def\Lm{\Lambda}

\def\pr{\prime}
\def\l{\left}
\def\r{\right}
\def\t{\tilde}

\def\Tr{{\rm Tr}}
\def\tr{{\rm tr}}

\begin{document}
\preprint{KAIST-TH 2003/06}
\title{\large \bf Flavor Hierarchy from Extra Dimension
and Gauge Threshold Correction}
\author{Kiwoon Choi\footnote{kchoi@hep.kaist.ac.kr}, Ian-Woo
  Kim\footnote{iwkim@hep.kaist.ac.kr} and Wan Young
  Song\footnote{wysong@hep.kaist.ac.kr}}
\address{Department of Physics, Korea Advanced Institute of Science and Technology\\ Daejeon
305-701, Korea}
\date{\today}

\begin{abstract}
Dynamical quasi-localization of matter fields
in extra dimension is an elegant mechanism
to generate hierarchical 4-dimensional Yukawa couplings.
We point out that a bulk matter field whose zero mode is 
quasi-localized can give a large Kaluza-Klein
threshold correction to low energy gauge couplings,
which is generically of the order of $\ln (\mbox{Yukawa})/8\pi^2$,
so  it can significantly affect gauge coupling unification.
We compute such threshold corrections 
in generic 5-dimensional theories compactified
on $S^1/Z_2\times Z_2$,
and apply the result
to grand unified theories on orbifold generating
small Yukawa couplings through 
quasi-localization.

\end{abstract}
\maketitle

\section{Introduction}

It has been noticed that extra dimension can provide
an elegant mechanism to 
generate hierarchical Yukawa
couplings
\cite{Arkani-Hamed:1999dc,Mirabelli:1999ks,Dvali:2000ha,Kaplan:2000av,Arkani-Hamed:2001tb,Kaplan:2001ga,Kakizaki:2001ue,Haba:2002uw,Hebecker:2002re,Grossman:2002pb,Chang:2002ww,Kitano:2003cn,Choi:2003di,Biggio:2003kp}.
The quark and lepton fields can be quasi-localized
in extra dimension in a natural manner, and then 
their 4-dimensional (4D) Yukawa couplings are determined by
the wavefunction overlap factor 
$e^{-M\pi R}$ where
$M$ is a combination of mass parameters in higher dimensional
theory and $R$ is the length of extra dimension.
This allows that hierarchical Yukawa couplings are obtained from
fundamental mass parameters having the same order of magnitude.
Extra dimension has been known to be useful also for
constructing a natural model of gauge unification 
\cite{Kawamura:2000ev,Altarelli:2001qj,Hall:2001pg,Hebecker:2001wq,Hebecker:2001jb,Hall:2001tn,Asaka:2001eh,Hall:2001xr,Dermisek:2001hp,Kim:2002ab,Kim:2002im}.
Supersymmetric 4D grand unified theories (GUTs) successfully
accomodate gauge coupling unification, however suffer from some
difficulties such as the doublet-triplet splitting problem and the proton
decay problem. 
These problems can be elegantly solved 
in GUT on orbifold with gauge symmetry 
broken by boundary conditions \cite{Kawamura:2000ev,Hebecker:2001wq}.
It is straightforward to implement the idea of
dynamical quasi-localization in orbifold GUT 
to get hierarchical Yukawa couplings as well as successful
gauge unification \cite{Kakizaki:2001ue,Hebecker:2002re}.

\medskip 

In any GUT, heavy particle threshold effects at
GUT symmetry breaking scale 
should be taken into account for a precision analysis of
low energy gauge couplings $g_a^2$.
In conventional 4D GUT,  those threshold corrections 
to $1/g_a^2$ are generically of the order of $1/8\pi^2$
and thus {\it not} so important,  {\it unless} 
the model contains a large number of
superheavy particles which become massive as a consequence
of GUT symmetry breaking 
 or some of superheavy masses are hierarchically different
from each other \cite{Weinberg:1980wa}.
As was pointed out a long time ago,
higher dimensional field theory and/or string theory
contain (infinitely) many Kaluza-Klein (KK) or stringy
modes, so can have a sizable threshold correction \cite{Choi:1987iu}.
Therefore it is essential to include stringy and/or KK threshold
correction in the precision analysis 
of low energy couplings in string and/or higher dimensional
field theories.
In this paper, we wish to examine the KK threshold corrections
in generic 5D orbifold field theories
in which hierarchical 4D Yukawa couplings are generated
through quasi-localization.
As we will see, the KK threshold corrections
to $1/g_a^2$ in such models are generically of the order of
$\ln (\mbox{Yukawa}) /8\pi^2$, so can significantly affect gauge 
coupling unification.
The outline of this paper is as follows.
In Sec. II, we discuss the KK spectrum and zero mode
wave functions of generic scalar and spinor fields in
5D theory compactified on $S^1/Z_2\times Z^\pr_2$.
In Sec. III, we discuss the 4D Yukawa couplings of
those quasi-localized scalar and fermion zero modes.
In Sec. IV, we  compute KK threshold corrections 
in generic 5D theories on $S^1/Z_2\times
Z^\pr_2$.
In Sec. V, we use these results 
to derive the KK  correction
to the {\it predicted} value of the low energy QCD coupling constant
in 5D GUT on orbifold.
We then construct a class of supersymmetric
5D orbifold GUTs which generate hierarchical Yukawa couplings
through dynamical quasi-localization while keeping
successful gauge coupling unification.
Sec. VI is the conclusion.

\section{Kaluza-Klein analysis and quasi-localized
zero modes}

In this section, we analyze the
KK wave functions and spectrums of scalar and spinor fields in 5D theory
compactified on $S^1/Z_2\times Z^\pr_2$.
Our major concern is the dynamical quasi-localization of
zero mode wavefunctions which would result in
hierarchical 4D Yukawa couplings.
We also consider the full KK spectrums which will be relevant 
for the discussion of KK threshold corrections to low energy gauge couplings.
Throughout this paper, we will use the convention for 
$S^1/Z_2\times Z_2^\pr$ which is given by the following transformation of 
the fifth spacetime coordinate $y\equiv y+4\pi R$: 
\bea
 Z_2 : y \rightarrow -y \,,
\quad\quad
 Z_2' : y+\pi R \rightarrow 
-y+\pi R\,. 
\eea
so the fundamental domain of $S^1/Z_2\times Z_2^\prime$ 
corresponds to $0\leq y\leq \pi R$.

\medskip

Let us first consider a 5D complex
scalar field with an action: 
\beq
-\int d^5x \,\left[\,
D_M\phi^{zz^\prime} D^M\phi^{zz^\prime*}+
\left(\,m_{zz^\prime}^2+2\mu_{zz^\prime}\delta(y)-
2\mu_{zz^\prime}^\prime\delta(y-\pi R)\,\right)
\phi^{zz^\prime}\phi^{zz^\prime*}\,\right]\,,
\label{scalaraction}
\eeq
and the orbifold boundary conditions:
\bea
&& \phi^{zz^\prime}(-y)=z\phi^{zz^\prime}(y)\,,
\nonumber \\
&& \phi^{zz^\prime}(-y+\pi R)=z^\prime\phi^{zz^\prime}
(y+\pi R)\,,
\nonumber 
\eea
where  $z=\pm 1$ and  $z^\prime=\pm 1$.
The equation of motion for each KK mode of this scalar field is given by
\beq
\label{scalarequation}
\left(\partial_y^2 +(\omega_n^{zz^\prime})^2\right)\phi^{zz^\prime}_n
=2\left[\,\mu_{zz^\prime}\delta(y)-
\mu^\prime_{zz^\prime}\delta(y-\pi R)\,\right]\phi^{zz^\prime}_n\,,
\eeq
where
\beq
\left(\omega_n^{zz^\prime}\right)^2=\left(M^{zz^\prime}_n\right)^2-
m^2_{zz^\prime}
\eeq
for $M_n^{zz^\prime}$ denoting the 4D mass of $\phi^{zz^\prime}_n$.
Upon ignoring unimportant normalization factor, one easily finds
the following KK wave functions on the fundamental domain:
\bea
&&\phi^{++}_n=\cos \left(\omega_n^{++} y\right)
+\frac{\mu_{++}}{\omega^{++}_n}\sin \left(\omega^{++}_n y\right) \,,
\nonumber \\
&& \phi^{+-}_n=\sin \left(\omega_n^{+-} (y-\pi R)\right) \,,
\nonumber \\
&& \phi^{-+}_n=\sin \left(\omega_n^{-+} y\right) \,,
\nonumber \\
&& \phi^{--}_n=\sin \left(\omega_n^{--} y\right) \,,
\eea
where $\omega_n^{zz^\prime} ( \neq 0 )$ are either real or
pure imaginary constants determined by
\bea
\tan \left(\omega_n^{++}\pi R\right)&=&\frac{(\mu_{++}-
\mu^\prime_{++})\omega_n^{++}}{
\left(\omega^{++}_n\right)^2+\mu_{++}\mu_{++}^\prime}\,, \nonumber \\
\tan\left(\omega_n^{+-}\pi R\right)&=&
-\frac{\omega^{+-}_n}{\mu_{+-}} \,, \nonumber \\
\tan\left(\omega_n^{-+}\pi R\right)&=&
\frac{\omega^{-+}_n}{\mu^\pr_{-+}}\,,
\nonumber \\
\tan\l(\omega_n^{--}\pi R\r)&=& 0. \label{massspectrum}
\eea
Once $\omega_n^{zz^\pr}$ are determined, the KK mass spectrum of
$\phi^{zz^\pr}$  can be determined by the relation
\beq
\l(M_n^{zz^\pr}\r)^2=m^2_{zz^\pr}+
\l(\omega_n^{zz^\pr}\r)^2.
\eeq
Note that a pure imaginary $\omega^{zz^\prime}_n$
gives  $\left(M_n^{zz^\prime}\right)^2$ smaller than $m_{zz^\pr}^2$,
while a real $\omega_n^{zz^\pr}$ gives $\l(M_n^{zz^\pr}\r)^2$
bigger than $m^2_{zz^\pr}$.
Note also that there can be
{\it only one or no} imaginary $\omega^{zz'}_n$  from
(\ref{massspectrum}).

\medskip

Since we  are interested in the shape of zero mode wavefunction,
let us consider the conditions for $\phi^{zz^\prime}$ to have
a zero mode, i.e $M^{zz^\prime}_0=0$,
{\it independently of} the value of the orbifold
radius $R$.
Obviously only $\phi^{++}$ can have such a zero mode
only when its bulk and brane masses are tuned to satisfy
\beq
\mu_{++}=\mu^\prime_{++}\,,
\quad
m^2_{++}=\mu^2_{++}\,,
\eeq
which are indeed satisfied in supersymmetric theories \cite{Arkani-Hamed:2001tb,Gherghetta:2000qt}.
The resulting zero mode (on the fundamental domain) 
is given by
\beq
\phi^{++}_0=\exp (\mu_{++}y )\,,
\eeq
so it is quasi-localized at $y=0$\, if\, $\mu_{++}<0$,
while at $y=\pi R$ \, if\, $\mu_{++}>0$.
In a  more general case with $\mu_{++}=
\mu^\pr_{++}$ but $m^2_{++}\neq \mu_{++}^2$,
the KK spectrums of $\phi^{++}$ are given by
\bea
&& \left(M^{++}_0\right)^2=m^2_{++}-\mu^2_{++}\,,
\nonumber \\
&& \left(M^{++}_n\right)^2=m^2_{++}+\frac{n^2}{R^2}\,,
\eea
for $n$ being a positive integer.

\medskip

For $\phi^{+-}$,\, if\,
$\mu_{+-}\pi R<-1$,  
there can be a mode
lighter than $m_{+-}$:
\bea
&& \phi^{+-}_0=\sinh \xi (y-\pi R) \,,
\nonumber \\
&& \left(M^{+-}_0\right)^2
=m_{+-}^2-\xi^2\,,
\eea
where $\xi$ is a real constant determined by
\bea
\tanh (\xi \pi R)=\frac{\xi}{|\mu_{+-}|}\,.
\nonumber
\eea
In the limit 
$\mu_{+-}\pi R \ll -1$,\,
$M^{+-}_0$ is determined to be
\beq
\left(M_0^{+-}\right)^2=\left(
m^2_{+-}-\mu^2_{+-}\right)
+4\mu^2_{+-}\left(
e^{-2|\mu_{+-}|\pi R}+
{\cal O}(e^{-4|\mu_{+-}|\pi R})\right).
\eeq
So in supersymmetric case in which
$m^2_{+-}=\mu^2_{+-}$,
the mass of $\phi^{+-}_0$ is exponentially suppressed
in the limit $\mu_{+-}\pi R \ll -1$.
On the other hand, all other KK modes of $\phi^{+-}$ are
heavier than  $m_{+-}$.
Similarly, \, if \, $\mu_{-+}\pi R> 1$, 
 $\phi^{-+}$ can have a lighter mode:
\bea
&& \phi^{-+}_0=\sinh (\xi^\pr y)\,,
\nonumber \\
&& \left(M^{-+}_0\r)^2=
m_{-+}^2-\xi^{\pr 2}\,,
\nonumber 
\eea
where 
$$
\tanh (\xi^\pr \pi R)=\frac{\xi^\pr}{\mu_{-+}}.
$$
In the limit $\mu_{-+}\pi R\gg 1$, we have
$$
\l(M^{-+}_0\r)^2=
\l(m^2_{-+}-\mu^2_{-+}\r)^2+
4\mu^2_{-+}\l(e^{-2\mu_{-+}\pi R}+
{\cal O}(e^{-4\mu_{-+}\pi R})\r),
$$
yielding an exponentially small mass of $\phi^{-+}_0$ 
in supersymmetric case with $\mu_{-+}\pi R\gg 1$.

\medskip

Let us now consider a 5D spinor field
with an action
\beq
\label{fermionaction}
-\int d^5x \,\l[\,i \bar{\Psi}^{zz^\prime}\left( \gamma^M \p_M + 
{\cal M}_{zz^\prime} \epsilon (y)
 \right) \Psi^{zz^\prime}\,\r]\,, 
\eeq
and the orbifold boundary condition
\bea
\Psi^{zz^\prime}(-y)&=&z\gamma_5\Psi^{zz^\pr}(y)\,,
\nonumber \\
\Psi^{zz^\prime}(-y+\pi R)&=&
z^\prime\gamma_5\Psi^{zz^\pr}(y+\pi R)\,,
\nonumber 
\eea
where $\epsilon(y)=-\epsilon(-y)=\epsilon(y+2\pi R)=1$ 
on the covering space of $S^1/Z_2\times Z^\prime_2$.
Note that in order for the action to be invariant under 
$Z_2\times Z^\prime_2$, 5D spinor  can have only a kink type mass
${\cal M}_{zz^\prime}\epsilon(y)$.
The equation of motion for each KK mode of $\Psi^{zz^\prime}$ 
is given by
\bea
&&\left(\partial_y^2+
(\omega_n^{zz^\prime})^2\right)^2 \Psi^{zz^\prime}_{nL} = 
-2{\cal M}_{zz^\pr}
\left[\,\delta(y)-\delta(y-\pi R)\,\right] \Psi^{zz^\prime}_{nL}
\,,
\nonumber \\
&&\left(\partial_y^2+
(\omega_n^{zz^\prime})^2\right)^2\Psi^{zz^\prime}_{nR} =
+2{\cal M}_{zz^\pr}\left[\,\delta(y)-\delta(y-\pi R)\,\right]
\Psi^{zz^\prime}_{nR},
\eea
where $\gamma_5\Psi_{L,R}=\pm \Psi_{L,R}$ and
\beq
\l(\omega_n^{zz^\pr}\r)^2=\l(M_n^{zz\pr}\r)^2-
\l({\cal M}^{zz^\pr}\r)^2
\nonumber
\eeq
for $M^{zz^\pr}_n$ denoting the 4D mass of $\Psi_n^{zz^\pr}$.
When compared to the scalar equation of motion 
(\ref{scalarequation}),
this suggests that in supersymmetric case the bulk and brane 
scalar masses should satisfy 
$$\mu_{zz^\prime}
=\mu^\prime_{zz^\prime},
\quad 
\mu^2_{zz^\prime} = m_{zz'}^2
={\cal M}_{zz^\prime}^2.
$$
Obviously,
 $\Psi^{++}$ has a left-handed
zero mode for any value of ${\cal M}_{++}$:
\beq
\Psi^{++}_{0L} (y )=\exp ( -{\cal M}_{++} y)
\eeq
which is quasi-localized at either $y=0$ (if ${\cal M}_{++}>0$) or $y=\pi R$
(if ${\cal M}_{++}<0$).
All other left-handed modes of $\Psi^{++}$
are paired-up with right-handed modes to 
get a 4D Dirac mass bigger than ${\cal M}_{++}$:
\beq
M^{++}_n = \sqrt{{\cal M}_{++}^2 + \f{n^2}{R^2}}\,.  
\eeq
Similarly, $\Psi^{--}$ has a right-handed zero mode
for any value of ${\cal M}_{--}$ and also the massive modes
with $(M_n^{--})^2={\cal M}^2_{--}+n^2/R^2$.
The wave function of zero mode is given by
\beq
\Psi^{--}_{0R} (y) =\exp ({\cal M}_{--}y), 
\eeq 
so the zero mode is quasi-localized at either $y=0$
(if ${\cal M}_{--}<0$) or $y=\pi R$ 
(if ${\cal M}_{--}>0$).

\medskip

For $\Psi^{+-}$,
there is no zero mode.
However 
if ${\cal M}_{+-}\pi R>1$,
there can be two modes lighter than ${\cal M}_{+-}$, while
all other modes are heavier than
${\cal M}_{+-}$ 
The wavefunctions of light modes
are  given by  
\bea
\Psi^{+-}_{0L} (y) &=& \sinh (k(y - \pi R ) ), \\
\Psi^{+-}_{0R} (y) &=& \sinh (k y ), \\
\eea
where $k$ is a real constant determined by
\beq
\tanh ({k\pi R}) = \frac{k}{{\cal M}_{+-}}.
\eeq
Obviously, one of these two light modes is localized
at $y=0$, while the other is localized at $y=\pi R$. 
In 4D  viewpoint, these two modes are paired up to get a Dirac mass
\beq
\l(M^{+-}_0\r)^2={\cal M}_{+-}^2-k^2\,.
\eeq
Note that in the limit ${\cal M}_{+-}\pi R\gg 1$,
\beq
M^{+-}_0 \approx 2{\cal M}_{+-}\exp (- {\cal M}_{+-} \pi R ),
\eeq
so $\Psi^{+-}_{0}$ can be arbitrarily light.
Similarly, if ${\cal M}_{-+}\pi R<-1$,
$\Psi^{-+}$ can have a light Dirac mode with
$M^{-+}_0\approx 2|{\cal M}_{-+}|\exp(
{\cal M}_{-+}\pi R)$.
\medskip

The dynamical quasi-localization of light modes and also
the shape of full KK spectrums offer a possibility that  
massless brane fields can be considered as a large
mass limit of bulk fields 
\footnote{Here we are implicitly assuming that there is a 3-brane at each
orbifold fixed point.}.
To see this, let us take the limit $|{\cal M}_{zz^\pr}|
\to \infty$ \footnote{This infinite kink mass should be
considered as a large mass comparable to the cutoff scale
of the orbifold field theory under consideration.}.
In this limit, 
the left-handed zero mode of $\Psi^{++}$
becomes a chiral brane fermion 
confined at $y=0$ (if ${\cal M}_{++}\rightarrow \infty$)
or at $y=\pi R$ (if ${\cal M}_{++}\rightarrow -\infty$),
while the right-handed zero mode of $\Psi^{--}$
becomes a chiral brane fermion at $y=0$ (if ${\cal M}_{--}\rightarrow
-\infty$) or at $y=\pi R$ (if ${\cal M}_{--}
\rightarrow \infty$).
For $\Psi^{+-}$, the light modes
$\Psi^{+-}_{0L}$ and $\Psi^{+-}_{0R}$ become a massless
brane fermion at $y=0$ and $y=\pi R$, respectively,
in the limit ${\cal M}_{+-}\rightarrow \infty$.
Similarly, $\Psi^{-+}_{0L}$ and $\Psi^{-+}_{0R}$
become a massless brane fermion at $y=\pi R$ and $y=0$, respectively,
in the limit  ${\cal M}_{-+}\rightarrow -\infty$.
All other KK modes of $\Psi^{zz^\pr}$ are
heavier than $|{\cal M}_{zz^\pr}|$, 
and thus are decoupled in the limit $|{\cal M}_{zz^\pr}|\rightarrow
\infty$.
So any massless brane fermion can be considered as
a bulk fermion in the large kink mass limit.

\section{Yukawa hierarchy from quasi-localization}

To discuss flavor hierarchy arising from quasi-localization,
let us consider a generic 5D theory containing arbitrary
number of scalar and fermion fields:
\bea
S&=&-\int d^5x \l[ \, D_M\phi_ID^M\phi^*_I
+\l(m_I^2\delta_{IJ}+2\mu_{IJ}\delta(y)-
2\mu^\pr_{IJ}\delta(y-\pi R)\r)\phi_I\phi^*_J \r.
\nonumber \\
&& \l.\quad\quad\quad\quad +\,i \bar{\Psi}_A(\gamma^MD_M+{\cal M}_A\epsilon(y))\Psi_A
+{\cal L}_Y\,\r]\,.
\eea
Here ${\cal L}_Y$ stands for the Yukawa couplings between
$\phi_I$ and $\Psi_A$
and the orbifold boundary conditions are given by
\bea
\phi_I(-y)=z_I\phi(y), && \,\,
\phi_I(-y+\pi R)=z_I^\pr\phi_I(y+\pi R)\,,
\nonumber \\
\Psi_A(-y)=z_A\gamma_5\Psi_A(y), && \,\,
\Psi_A(-y+\pi R)=z_A^\pr \gamma_5 \Psi_A(y+\pi R)\,.
\eea
Though the brane masses of scalar fields
can have off-diagonal components in general, we will
assume for simplicity that they are diagonal:
$\mu_{IJ}=\mu_I\delta_{IJ}$ and
$\mu^\pr_{IJ}=\mu^\pr_I\delta_{IJ}$. 
To get hierarchical 4D Yukawa couplings, we also assume that
Yukawa couplings in 5D theory exist  only at the
fixed points, which is assured by 5D SUSY in supersymmetric
theories.
Then the most general form of Yukawa couplings can be written as
\beq
 {\cal L}_Y =
\delta(y)\, \f{{\lambda}_{_{XPQ}}}{\Lambda^{3/2}}\,
 \varphi_{_X} \psi_{_P}\psi_{_Q}
+\delta(y-\pi R)\,\f{{\lambda}^\pr_{_{XPQ}}}{\Lambda^{3/2}}\,
\varphi_{_X}\psi_{_P}\psi_{_Q}
+{c.c.},
\eeq
where $\Lambda$ denotes the cutoff scale of our 
5D orbifold field theory,
$\varphi_{_X}=\{\phi_I \,\,\, \mbox{or} \,\,\, \phi^*_I\}$,
and $\psi_{_P}=
\{ \f{1}{2}(1+\gamma_5)\Psi_{A} \,\, \mbox{or} 
\,\, \f{1}{2}(1+\gamma_5)\Psi_{A}^c\}$.
Note that 
$\lambda_{ipq}$ and $\lambda^\pr_{ipq}$ are dimensionless
parameters in our convention.

\medskip

Let $\phi_i$ denote 5D scalar fields having a zero mode
$\phi_{0i}=\exp (m_i y)$,
i.e. scalar fields with $z_i=z^\pr_i=1$ and $\mu_i=\mu^\pr_i=m_i$,
and $\Psi_p$ denote 5D spinor fields having a chiral zero mode
$\psi_{0p}=\exp (- z_p {\cal M}_p y)$, i.e.
spinor fields with $z_p=z^\pr_p=\pm 1$.
It is then straightforward to find that the 4D Yukawa couplings
of {\it canonocally normalized} zero modes are given by
\bea
\label{canonicalyukawa}
y_{ipq} 
&=& 
\sqrt{Z(m_i)Z(-z_p{\cal M}_p)Z(-z_q{\cal M}_q)}\,\,\lambda_{ipq}
\nonumber \\
&&+\sqrt{Z(-m_i)Z(z_p{\cal M}_p)Z(z_q{\cal M}_q)}\,\,
\lambda^\pr_{ipq}
\eea
where 
\bea
Z(m) =\f{2m}{\Lambda}\f{1}{e^{2m\pi R}-1}.
\eea
Obviously, $y_{ipq}$ can have very different values,
depending upon the values of $m_i$ and ${\cal M}_{p,q}$,
even when all of the 5D parameters $\lambda_{ipq}$ and $\lambda^\pr_{ipq}$
have similar values.

\medskip

A simple way to get hierarchical Yukawa couplings
through quasi-localization is to assume that all Yukawa couplings 
originate from a single fixed point, for instance from $y=\pi R$.
As a concrete example, let us consider the case that
all Yukawa couplings originate from $y=\pi R$ and
$\phi_i$ is a brane field confined at
$y=\pi R$ with $Z(-m_i)\approx 1$. 
Note that a brane scalar field at $y=\pi R$  can be obtained from
a bulk scalar field by taking the limit  
$m={\cal O}(\Lambda)$ while keeping $\mu=
\mu^\pr=m$.
Then for $z_{p,q}{\cal M}_{p,q}\lesssim - 1/R$,
we have
\beq
y_{ipq}\approx \sqrt{\f{4|{\cal M}_p{\cal M}_q|}{\Lambda^2}}
\,\,{\lambda}^\pr_{ipq}\,,
\eeq
while for $z_{p,q}{\cal M}_{p,q}\gtrsim 1/R$,
\beq
\label{smallyukawa}
y_{ipq}\approx \sqrt{\f{4|{\cal M}_p{\cal M}_q|}{\Lambda^2}}\,
e^{-(z_p{\cal M}_p+z_q{\cal M}_q)\pi R}\,\,{\lambda}^\pr_{ipq}.
\eeq
The physical interpretation of this result is simple.
If $z_{p,q}{\cal M}_{p,q}\gtrsim 1/R$, 
the corresponding zero modes are quasi-localized at  $y=0$, 
so the Yukawa couplings are exponentially 
suppressed as they originate from $y=\pi R$.
On the other hand, for $z_{p,q}{\cal M}_{p,q}\lesssim
-1/R$, the zero modes are localized at $y=\pi R$, so
there is no exponential suppression in Yukawa couplings.

\medskip 

If the fundamental theory at $\Lambda$
is weakly coupled, simple dimensional analysis
would suggest that the dimensionless 
${\lambda}_{ipq}$ are of order unity or less. 
However a more interesting possibility is that the theory
is {\it strongly} coupled at $\Lambda$ \cite{Chacko:1999hg,Nomura:2001tn}.
In 5D theories under consideration, the standard model gauge fields 
live in bulk spacetime, so their 4D couplings
are given by 
$1/{g_a^2}\approx {\pi R}/{g_{5a}^2}$,
where $g^2_{5a}$ denote the dimensionful 5D gauge couplings with mass
dimension $-1$.
In order for our 5D theory to be a useful framework,
it must be valid up to an energy scale significantly 
higher than the compactification scale $1/R$, i.e. $\Lambda\gg 1/R$.
If $\Lambda$ were comparable to $1/R$, we would not have to consider 
a 5D theory as an intermediate step going
from the fundamental theory at $\Lambda$ to the 4D effective
theory for low energy physics.  We would rather go directly to 
the 4D effective theory from the fundamental theory.
On the other hand, if $\Lambda\gg 1/R$ as desired,
we have  
$g_{5a}^2\approx \pi R g_a^2 \gg {1}/{\Lambda}$,
implying that the fundamental theory at $\Lambda$ is strongly coupled.
As is well known, only in strongly coupled scenario,
GUT on orbifold can provide a meaningful 
prediction for $\sin^2\theta_W$ or the QCD coupling constant
at the weak scale.
In such strongly-coupled scenario,  dimensional analysis suggests
$$
{\lambda}^\pr_{ipq}={\cal O}(4\pi).
$$
It is then straightforward to get hierarchical 4D Yukawa couplings
ranging from  the top quark Yukawa coupling 
$y_t\approx 1$ to the electron Yukawa coupling $y_e\approx
10^{-5}-10^{-6}$ within the parameter range
$$
\Lambda\pi R={\cal O}(10^2)\,,\quad
|{\cal M}_p|\pi R\lesssim 7\,.
$$
Note that, in strongly coupled scenario,
if the Higgs boson and the left and
right-handed top quarks are all brane fields,
the resulting top-quark Yukawa coupling 
would be too large, $y_t={\cal O}(4\pi)$, so
one needs to put some of those fields in bulk
spacetime.

\medskip

In supersymmetric 5D theories, 
the fermion kink masses ${\cal M}_p$ are related to 
the  graviphoton gauging \cite{Ceresole:2000jd,Choi:2002wx}:
\beq
D_M\Psi_p=\partial_M\Psi_p+i{\cal M}_p\epsilon(y)B_M\Psi_p+...\,,
\eeq
where $B_M$ denotes the graviphoton. 
This suggests that it is a plausible assumption that
${\cal M}_p$ are 
{\it quantized} (in an appropriate unit) as the
conventional gauge charges are quantized.
If true, the resulting Yukawa couplings
(\ref{smallyukawa}) would have the same form  
as those obtained from the
Frogatt-Nielson mechanism which
generates small Yukawa couplings using
a spontaneously broken $U(1)$ flavor symmetry \cite{Froggatt:1978nt}.
In the next section, we will see that 
the KK threshold corrections to low energy gauge couplings
are generically given by
\beq
\Delta\l(\f{1}{g_a^2}\r)={\cal O}\l(
\f{\ln (y_{ipq})}{8\pi ^2}\r),
\eeq
if the hierarchical 4D Yukawa
couplings $y_{ipq}$ are generated by quasi-localization.
Obviously then the KK threshold corrections
can significantly affect the gauge coupling unification.

\medskip

\section{kaluza-klein threshold correction to low energy gauge coupling}

In this section, we discuss the 1-loop threshold correction
to low energy gauge coupling 
in 5D orbifold field theory.
The bare action of bulk gauge fields 
can be written as
\bea
S_{\rm bare} = -\int d^5x \l(\f{1}{4g_{5a}^2}
+\f{\kappa_a}{4}\delta(y)
+\f{\kappa^\pr_a}{4}\delta(y-\pi R)\r)
F^{aMN}F^a_{MN}, \label{bareaction}
\eea
and then the 4D gauge couplings at tree level are given by
\beq
\l(\f{1}{g_a^2}\r)_{\rm tree}=\f{\pi R}{g_{5a}^2}
+\kappa_a+\kappa^\pr_a.
\eeq
One-loop corrections to low energy couplings can be
computed by summing the contributions
from  all KK modes. In generic 5D orbifold field theory,
such computation yields \cite{Contino:2001si,Choi:2002zi}  
\bea
\label{lowenergycoupling}
\f{1}{g_a^2(p)}&=&
\l[\f{\pi R}{g_{5a}^2}+\f{\gamma_a}{24\pi^3}\Lambda\pi R+
\kappa_a+\kappa^\pr_a\r]
+\f{1}{8\pi^2}\l[{\Delta}_a(\ln\Lambda, m, \mu,
\mu^\pr, {\cal M}, R)+
b_a\ln\l(\f{\Lambda}{p}\r)\r]\,
\nonumber \\
&\equiv& \l(\f{1}{g_a^2}\r)_{\rm bare}+
\f{1}{8\pi^2}\l[\Delta_a(\ln\Lambda, m, \mu,
\mu^\pr, {\cal M}, R)+b_a\ln\l(\f{\Lambda}{p}\r)\r]
\eea
where $\gamma_a$ are the coefficients of 
(UV-sensitive) linearly divergent corrections,
$\Delta_a$ stand for (UV-insensitive) logarithmically divergent or finite
threshold corrections due to massive KK modes, and 
$b_a$ are the standard 1-loop beta function
coefficients due to zero modes.
Here $(1/g_a^2)_{\rm bare}$ correspond to 
the {\it uncalculable} bare couplings of  the model,
while $\Delta_a$ are unambiguously calculable
within 5D orbifold field theory.
Note that $\Delta_a$ contain a piece linear in $\ln\Lambda$
as well as a finite piece depending on the scalar and fermion 
mass parameters $m,\mu,\mu^\pr,{\cal M}$ and also on 
the orbifold radius $R$.
Here we assume that the cutoff scale $\Lambda$ is large enough compared
to other mass parameters of the theory, thus ignore the part
suppressed by an inverse power of $\Lambda$.
In the above expression, the renormalization point 
$p$ is assumed to be below the mass of the {\it lightest
massive} KK mode, $M_{KK}$, but
far above the masses of all zero modes which would be
around the weak scale.

\medskip

The KK threshold corrections in 5D theories
on {\it warped} $S^1/Z_2\times Z^\pr_2$ have been discussed before
in \cite{Choi:2002wx,Choi:2002zi,Choi:2002ps}.
In \cite{Choi:2002wx,Choi:2002zi}, $\Delta_a$ for
supersymmetric 5D theories on warped 
$S^1/Z_2\times Z^\pr_2$ have been computed  
in the framework of 4D effective supergravity.
In this framework, $\Delta_a$ could be obtained 
by computing the tree-level K\"ahler potential
and also the {\it one-loop} correction
to holomorphic gauge kinetic functions
which can be determined by the chiral anomaly structure
of 5D orbifold field theory \cite{Arkani-Hamed:2001is}.
The KK threshold corrections for nonsupersymmetric
5D theories on warped $S^1/Z_2\times Z^\prime_2$
have been computed in \cite{Choi:2002ps} by directly evaluating
all KK mode contributions
in dimensional regularization scheme \cite{GrootNibbelink:2001bx}, 
and it was confirmed that
the results in supersymmetric limit agree with 
those of \cite{Choi:2002wx,Choi:2002zi}.
In this paper, we compute the KK threshold corrections
in generic 5D theories on {\it flat} $S^1/Z_2\times Z_2^\pr$
with quasi-localized zero modes using the method of
\cite{Choi:2002ps}.

\medskip

To calculate one-loop gauge couplings at low energies, we
integrate out all massive KK modes and derive the one-loop effective
action of gauge field zero mode $A_\mu$: 
\beq
S_{\rm eff} = S_{\rm bare} + \Gamma_\phi  +
         \Gamma_\Psi  + \Gamma_A \,, 
\eeq
where $S_{\rm bare}$ is given in (\ref{bareaction}), 
and the 1-loop contributions from 5D scalar $\phi$,
5D spinor $\Psi$ and
5D vector $A_M$  are given by
\bea
i \Gamma_\phi  &\,=\,& -\f{1}{2} \Tr_\phi \ln \,( - D^2 +
M^2(\phi) )\,, \nonumber \\
i \Gamma_\Psi  &\,=\,& \f{1}{2} \Tr_\Psi \ln \,( -D^2 + M^2(\Psi)
+ F_{\mu\nu} J^{\mu\nu}_{1/2}) 
+ \Tr_{\Psi_0} \ln  (\,D \!\!\!\! /\,)\,,   
\nonumber \\
i \Gamma_A  &\,=\,& -\f{1}{2} \Tr_{A_\mu} \ln \,( -D^2 + M^2
(A_\mu) +F_{\mu\nu} J^{\mu\nu}_1)  - \f{1}{2} \Tr_{A_5} \ln \,(-D^2 +M^2
(A_5) ) \nonumber \\
&&+ \,\Tr_{\xi \bar{\xi} } \ln \,( - D^2 +M^2(\xi) )\,. 
\eea
Here $\Tr_\Phi$ implies the functional trace for the
5D field $\Phi$, so contains the summations over the whole KK modes.
$M^2 (\Phi)$ denotes the mass-square operator whose
eigenvalues correspond to the KK mass spectrum of $\Phi$, and
$J^{\mu\nu}_j$ is the 4D Lorentz generator for spin $j$, normalized 
as $\tr (J_j^{\mu\nu} J_j^{\rho\sigma} )= C(j) (g^{\mu\rho} g^{\nu\sigma} -
g^{\mu\sigma} g^{\nu\rho} ) 
$ 
where $C(j) = ( 0 ,1 , 2)$ for $j = (0 , 1/2 , 2)$. $\Tr_{\Psi_0}$
denotes the functional trace for the zero mode of $\Psi$ and
$\Tr_{\xi \bar{\xi}}$ is the trace over the ghost
fields.
   
\medskip

The most convenient way to calculate ${\rm Tr}_{\Phi}$ is to
replace the KK summation by a
contour integration with an appropriately chosen pole function $P(q)$.
The pole function we will use here has the form 
\beq 
P(q) = \f{N'(q)}{2 N(q) }\,,
\eeq
where $N(q)$ has zeroes at $ q^2 = M^2_n (\Phi) - m_{\Phi}^2$ for  
 $M_n (\Phi)$ denoting the n-th KK mass and $m_\Phi$ denoting
 the bulk mass of $\Phi$. Note that our pole function is different 
from the pole function of \cite{GrootNibbelink:2001bx,Choi:2002ps}
as the pole positions are shifted by $m^2_\Phi$, which is mainly
for the simplicity of calculation.
Using the analysis of Sec. II, we find the following forms of
$N$-functions for 5D scalar and fermion fields: 
\bea
N_{\phi^{++}}(q) &=& -\f{1}{q} ( q^2 + \mu_{++} \mu'_{++} ) \sin (q\pi R)
+ (\mu_{++} - \mu'_{++} ) \cos (q\pi R)\,, \nonumber \\
N_{\phi^{+-}}(q) &=& \f{1}{q} (\mu_{+-}\sin (q\pi R) + q \cos (q\pi R))\,,
\nonumber \\
N_{\phi^{-+}}(q) &=& \f{1}{q} (-\mu'_{-+} \sin (q\pi R) + q \cos (q\pi
R))\,,
  \nonumber \\
N_{\phi^{--}}(q) &=& \f{1}{q} \sin (q\pi R)\,, \nonumber \\
N_{\Psi^{++}} (q) &=& \f{1}{q} \sin (q\pi R)\,, \nonumber \\
N_{\Psi^{+-}} (q) &=& \f{1}{q} ( -{\cal M}_{+-} \sin (q\pi R) + q \cos
(q\pi R))\,, \nonumber \\
N_{\Psi^{-+}}(q) &=& \f{1}{q} ( {\cal M}_{-+} \sin (q\pi R ) + q \cos
(q\pi R) )\,, \nonumber \\
N_{\Psi^{--}} (q) &=& \f{1}{q} \sin{q \pi R}\,, 
\eea
where the scalar and fermion mass parameters 
$\mu_{zz^\pr}$, $\mu^\pr_{zz^\pr}$ and
${\cal M}_{zz^\pr}$ are defined in
(\ref{scalaraction}) and (\ref{fermionaction}). 
The $N$-functions for 5D vector fields $A_M^{zz^\pr}$ are also easily
found to be
\bea
N_{A^{++}} (q) &=& \f{1}{q} \sin (q\pi R)\,, \nonumber \\
N_{A^{+-}} (q) &=& \cos (q \pi R)\,, \nonumber \\
N_{A^{-+}} (q) &=& \cos (q \pi R)\,, \nonumber \\
N_{A^{--}} (q) &=& \f{1}{q} \sin (q\pi R)\,, 
\eea
where the boundary conditions of $A^{zz^\pr}_M$ are given by
\bea
&& A^{zz^\pr}_\mu (-y)=zA^{zz^\pr}_\mu (y)\,,\quad 
A^{zz^\pr}_\mu (-y+\pi R)=z^\pr A^{zz^\pr}(y+\pi R)\,,
\nonumber \\
&&
A^{zz^\pr}_y (-y)=-zA^{zz^\pr}_y(y)\,,\quad
A^{zz^\pr}_y(-y+\pi R)=-z^\pr A^{zz^\pr}_y(y+\pi R)\,.
\nonumber 
\eea

\medskip

\begin{figure}
\centering
\caption{\label{contour1} Contour $C_1$ in the complex q-plane. Bold
  dots represent poles at $q^2 = M^2_n (\Phi) - m^2_\Phi$ where
$M_n$ is the KK mass eigenvalue and $m_\Phi$ is the bulk mass of
5D field $\Phi$. 
Note that a 4D state with $M_n< m_\Phi$ 
appears as a pole at the imaginary axis.} 
\epsfig{figure=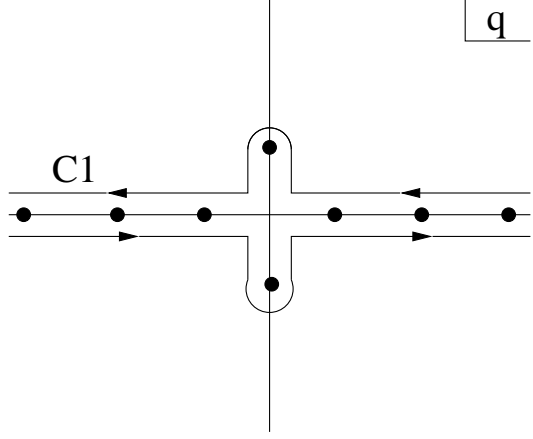,width=7cm}
\end{figure}
\begin{figure}
\centering
\caption{\label{contour2} The contour $C_1$ on the upper
  half-plane can be deformed into $C_2$ without touching any
  singularity. The poles at the imaginary axis do not overlap with
 the brance cut unless there is a tachyon state. }
\epsfig{figure=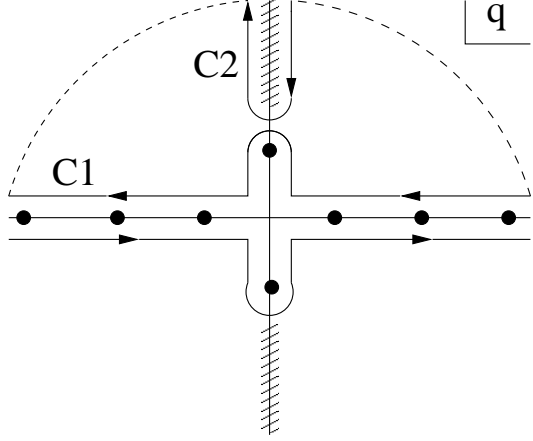,width=7cm}
\end{figure}

For the pole function defined as above, it is straightforward
to find
\bea
&&\Tr \ln \, ( -D^2 + M^2(\Phi) + F_{\mu\nu}J^{\mu\nu}_j ) \\
& = &\int_{C_1} \f{d q}{2\pi i }
\,P(q)\,
\int \f{d^4 p}{(2\pi)^4} A^a_\mu (-p) A^a_\nu (p) T_a(\Phi) \nonumber\\
&& \times \l[ d(j) \int \f{d^4 k}{(2\pi)^4} 
\f{g^{\mu\nu} \l((p+k)^2+q^2 +m_\Phi^2  \r) - \f{1}{2} (p+2k)^\mu (p+2k)^\nu }
{\l( k^2 +q^2 + m_\Phi^2 \r) \l( (p+k)^2 +q^2 + m_\Phi^2 \r)} \r. \nonumber \\
&& \quad\quad \l.
-2C(j)\l(p^2 g^{\mu\nu}-p^\mu p^\nu \r) 
\int \f{d^4 k}{(2\pi)^4} \f{1}{\l( k^2 +q^2 + m_\Phi^2 \r) \l( (p+k)^2
  +q^2 + m_\Phi^2  \r)}\,\r]
\nonumber \\
&\equiv& i \int \f{d^4 p}{(2\pi)^4} \,{\cal G}_a(p) A^a_\mu (-p)  
\l(p^2 g^{\mu\nu} - p^\mu p^\nu \r)A^a_\nu(p)\,,
\label{trlogd2}
\eea
where  
$d(j)=(1,4,4)$ and $C(j)=(0,1,2)$  for $j=(0,1/2,1)$,
and
 $T_a(X)=
{\rm Tr}\l(T^2_a(X)\r)$ is the Dynkin index of the gauge
group representation $X$. 
Here the contour $C_1$ is depicted in Fig. \ref{contour1}. 
To regulate the divergent part of the above integral, we split the 
pole function into two parts:
\beq
\label{decomposition}
P(q) = \tilde P (q) + P_\infty (q), 
\eeq
where $\tilde P \to O(q^{-2}) $  at $|q|\to \infty$.  
Then $P_{\infty}$ is given by
\beq
P_{\infty}(q) = -\f{A}{q} -\f{i\pi R}{2}\epsilon({\rm Im}(q) )\,,
\eeq
where $\epsilon(x)=x/|x|$ and
$A$ is a real constant depending on the 
$Z_2\times Z_2^{\prime}$ parity of the
corresponding 5D field:
$$A= \l(-1/2, \, 0,\, 0,\,1/2\r)$$ 
for $Z_2\times Z_2^{\prime}$ parity 
 $(Z_\Phi,Z^\pr_\Phi)=(++,+-,-+,--)$. 
With the decomposition (\ref{decomposition}), 
all UV divergences appear in
the contribution from $P_{\infty}$ in a manner allowing simple
dimensional regularization.

\medskip

The 4D momentum integral $d^4 k$ 
in (\ref{trlogd2}) exhibits 
a branch cut on the imaginary axis of $q$ for $p^2 > 0$. For the 
contribution from $\t{P}$, 
one can change the contour as in Fig. \ref{contour2} 
since the contribution from the infinite half-circle vanishes.
Note that the poles of $\t{P}$ on the imaginary axis do not overlap
with the branch cut as long as there is no tachyon.    
After integrating by part,  we find that
the part of ${\cal G}_a$ from $\t{P}$ is given by
\bea
\Delta {\cal G}_a(\t{P})&=&
\f{T_a(\Phi)}{8\pi^2} \l( \f{1}{6} d(j) - 2 C(j) \r)
{\cal F}(q)\Big|_{q\to i\infty}
\nonumber \\
&&-\f{T_a(\Phi)}{8\pi^2} \int_0^1 dx \l( \f{1}{2} d(j) (1-2x)^2 - 2 C(j) \r)
{\cal F}(q)\Big|_{q = i\sqrt{x(1-x) p^2 + m_\Phi^2}}\,,
\eea 
where 
$$
{\cal F}(q)=\f{1}{2} \ln N + A \ln q + \f{i \pi R}{2} q \,.
$$
In fact, ${\cal F}(q) \Big|_{q \to i\infty}$ turns out to be vanishing in the
cases we are now considering.  
The contribution from $P_{\infty}$ includes the log divergence
from the pole term $1/q$. This can be regulated by the standard 
dimensional regularization of 4D momentum integral, 
$d^4p\to d^Dp$, yielding
a $1/(D-4)$ pole. On the other hand, the step-function contribution
from $\epsilon({\rm Im}(q))$ involves a 5D momentum integral
which is linearly divergent, but it simply gives a finite result in 
dimensional regularization.
Note that linearly divergent correction to $1/g_a^2$
depends highly on the used regularization scheme.
Namely the coefficient $\gamma_a$ of Eq. (\ref{lowenergycoupling})
is regularization scheme dependent, and $\gamma_a=0$
in the dimensional regularization scheme we are currently using.
However this does not have any special meaning since the physical
amplitudes are always expressed in terms of
the scheme-independent combination $\f{\pi R}{g_{5a}^2}+
\f{\gamma_a}{24\pi^3}\Lambda \pi R$.
Adding the divergent contribution from ${P}_{\infty}$ to the finite part
$\Delta {\cal G}_a$ from $\t{P}$, we obtain 
\bea
\label{twopoint}
{\cal G}_a&=& \f{T_a(\Phi)}{8\pi^2}\l[ 
\int_0^1 dx \, \l(-\f{1}{2}d(j)(1-2x)^2+2C(j)\r)
 \l(\f{1}{2} \ln N  \r) \Big|_{q = i\sqrt{x(1-x) p^2 + m_\Phi^2}}
 \r. \nonumber \\
&& \l. 
+A \int_0^1 dx \l( -\f{1}{2} d(j) (1-2x)^2 +2C(j) \r) \l( \f{1}{D-4} \r)\r]\,.
\eea

Using the above result, we find that the KK
threshold corrections from a 5D complex scalar $\phi$, 5D Dirac fermion
$\Psi$ and 5D vector $A_M$ are given by
\bea 
\Delta_a(\phi) + b^\phi_a \ln \f{\Lm}{p} & =
&\f{1}{6}\l[ 
T_a(\phi^{++})\l\{\ln \Lm 
-3\int_0^1 du F(u)\ln N_{\phi^{++}}\l(i \sqrt{\f{u^2 p^2}{4} + m_{++}^2} \r)
\r\}
\r.
\nonumber \\
&&\quad \quad -\,3T_a (\phi^{+-})
\int_0^1 du\, F(u)\ln N_{\phi^{+-}} \l(i \sqrt{\f{u^2 p^2}{4} + m_{+-}^2}\r)
\nonumber \\
&&\quad\quad -\,3T_a (\phi^{-+})
\int_0^1 du\, F(u)\ln N_{\phi^{-+}} \l(i \sqrt{\f{u^2 p^2}{4} + m_{-+}^2}\r)
\nonumber \\
&&\l.\quad\quad -\,T_a (\phi^{--})  \l\{ \ln \Lm  
+3 \int_0^1 du\, F(u)\ln N_{\phi^{--}} \l(i \sqrt{\f{u^2 p^2}{4} + m_{--}^2}\r)
\r\}\,\r], 
\nonumber \\
\Delta_a(\Psi) + b_a^\Psi \ln \f{\Lm}{p} &=&
\f{1}{3}\l[
T_a (\Psi^{++})\l\{-2\ln p 
+3\int_0^1 du \,G(u)
\ln N_{\Psi^{++}} \l(i \sqrt{\f{u^2 p^2}{4} + {\cal M}_{++}^2}\r)
\r\}\r.\nonumber \\
&&\quad\quad +\,T_a (\Psi^{+-})\l\{   
+3\int_0^1 du \,G(u) 
\ln N_{\Psi^{+-}}\l(i \sqrt{\f{u^2 p^2}{4} + {\cal M}_{+-}^2}\r)
 \r\}\nonumber \\
&&\quad\quad +\,T_a (\Psi^{-+}) \l\{   
+3\int_0^1 du \,G(u)
\ln N_{\Psi^{-+}} \l(i \sqrt{\f{u^2 p^2}{4} + {\cal M}_{-+}^2}\r)
\r\}
\nonumber \\
&&\l.\quad\quad +\,T_a (\Psi^{--})\l\{-2\ln {p}
+3\int_0^1 du \, G(u)\ln N_{\Psi^{--}} \l(i \sqrt{\f{u^2 p^2}{4} +
  {\cal M}_{--}^2}\r)
\r\}\,\r]\,
\nonumber \\
\Delta_a(A_M) + b_a^A \ln \f{\Lm}{p} &=&
\f{1}{12}\l[
T_a(A^{++}_M)\l\{- 23\ln \Lm +44 \ln p
+\int_0^1 du \,K(u)\ln N_{A^{++}}\l(\f{iu}{2} \sqrt{p^2}\r)
\r\}\r.\nonumber \\
&&\quad\quad +\,T_a(A^{+-}_M) \int_0^1 du\, K(u)
\ln N_{A^{+-}}\l(\f{iu}{2} \sqrt{p^2}\r)
\nonumber \\
&&\quad\quad +\,T_a(A^{-+}_M) 
\int_0^1 du\, K(u) \ln N_{A^{-+}}\l(\f{iu}{2} \sqrt{p^2}\r)
\nonumber \\
&&\l.\quad\quad +\,T_a(A^{--}_M)\l\{ 23 \ln \Lm-2\ln p
+\int_0^1 du\, K(u)
\ln N_{A^{--}}\l(\f{iu}{2} \sqrt{p^2}\r)
\r\}\,\r]\,,
\nonumber 
\eea
where 
$b_a^\phi$,
$b_a^\psi$ and $b_a^A$ are the beta function coefficients for
the massless modes from $\phi$, $\Psi$ and $A_M$,
respectively, and
\bea
F(u) &=& u(1-u^2)^{1/2} \,,\nonumber \\
G(u) &=& u(1-u^2)^{1/2} - u(1-u^2)^{-1/2}\,, \nonumber\\
K(u) &=& -9u(1-u^2)^{1/2} +24 u(1-u^2)^{-1/2}\,. \nonumber
\eea 
For the case that $\sqrt{p^2}$ is much smaller than the lowest nonzero
KK mass, the above results are simplified to yield
\bea
\label{result1}
\Delta_a &&=\,
\frac{21}{12}\l[T_a(A^{++}_M)+T_a(A^{--}_M)\r]\ln(\Lambda\pi R)
\nonumber \\
&&-\f{1}{6}T_a({\Phi}^{++})\ln \l(\f{\Lambda(e^{m_{++}\pi R}
-e^{-m_{++} \pi R})}{2m_{++}}\r)
\nonumber \\
&&-\f{1}{6}T_a(\phi^{++})\ln\l(\f{(m_{++}+\mu_{++})(m_{++}-\mu^\pr_{++})e^{m_{++}\pi R}
-(m_{++}-\mu_{++})(m_{++}+\mu^\pr_{++})e^{-m_{++}\pi R}}{2m_{++}\Lambda}\r)
\nonumber \\
&&-\f{1}{6}T_a(\phi^{+-})\ln\l(
\f{(m_{+-}+\mu_{+-})e^{m_{+-}\pi R}+(m_{+-}-\mu_{+-})e^{-m_{+-}\pi R}}{2m_{+-}}\r)
\nonumber \\
&&-\f{1}{6}T_a(\phi^{-+})\ln\l(
\f{(m_{-+}-\mu^\pr_{-+})e^{m_{-+}\pi R}+(m_{-+}+\mu^\pr_{-+})e^{-m_{-+}\pi R}}{2m_{-+}}\r)
\nonumber \\
&&-\f{1}{6}T_a(\phi^{--})\ln\l(
\f{\Lambda (e^{m_{--}\pi R}-e^{-m_{--}\pi R})}{2m_{--}}\r)
\nonumber \\
&&-\f{2}{3}T_a(\Psi^{++})\ln\l(\f{\Lambda(e^{{\cal M}_{++}\pi R}
-e^{-{\cal M}_{++}\pi R})}{2{\cal M}_{++}}\r)
\nonumber \\
&&-\f{2}{3}T_a(\Psi^{+-})\ln \l(e^{-{\cal M}_{+-}\pi R}\r)
\nonumber \\
&&-
\f{2}{3}T_a(\Psi^{-+})\ln\l(e^{{\cal M}_{-+}\pi R}\r)
\nonumber \\
&&-\f{2}{3}T_a(\Psi^{--})\ln\l(
\f{\Lambda (e^{{\cal M}_{--}\pi R}-e^{-{\cal M}_{--}\pi R})}{2{\cal M}_{--}}\r)\,,
\eea
where $m_{zz'}, \mu_{zz'}$ and  $\mu^\pr_{zz'}$
denote the bulk and brane masses of $\phi$, and ${\cal M}_{zz'}$
is the kink mass of $\Psi$.
Here
${\Phi}^{++}$ is a 5D complex scalar field
having a zero mode, i.e. a scalar field with
$\mu=\mu^\pr=m$, and $\phi^{++}$ stands for complex scalar fields
{\it without} zero mode.
The 4D beta function coefficients $b_a$ are given by
\beq
b_a=-\f{11}{3}T_a(A^{++}_M)+\f{1}{6}T_a(A^{--}_M)
+\f{1}{3}T_a({\Phi}^{++})+\f{2}{3}T_a(\Psi^{++})
+\f{2}{3}T_a(\Psi^{--}),
\eeq
which can be easily understood by noting that
$A^{++}_M$ gives a massless 4D vector, $A^{--}_M$
a massless real 4D scalar, and $\Psi^{\pm\pm}$
a massless 4D chiral fermion.
Then comparing the above $\Delta_a$ to the expression of
4D Yukawa couplings in (\ref{canonicalyukawa}),  
one easily finds that generically
\beq
\Delta_a={\cal O}\l(\ln (y_{ipq})\r)
\eeq
if the 4D Yukawa couplings $y_{ipq}$ are
generated by quasi-localization.

\medskip

It is straightforward to find an expression of $\Delta_a$
for supersymmetric 5D theories using the above result.
In supersymmetric theories, there can be two type of bulk fields:
vector multiplet ${\cal V}$
containing a 5D vector $A_M$,  a Dirac-spinor 
$\lambda$  and a real scalar $\Sigma$,
and hypermultiplet ${\cal H}$
containing a Dirac spinor $\Psi$ and
two complex scalars $\phi,\phi^\pr$.
The mass parameters and $Z_2\times Z_2^\pr$ boundary conditions 
of component fields are given by
\bea
&&{\cal V}^{zz^\pr}=\{\,
A^{zz^\pr}_M,\, \lambda^{zz^\pr}({\cal M}=0), 
\,\Sigma^{\tilde{z}\tilde{z}^\pr}(\mu=\mu^\pr=m=0)\,\}\,,
\nonumber \\
&&{\cal H}^{zz^\pr}=\{\,
\phi^{zz^\pr}(\mu=\mu^\pr=m=-{\cal M}), \,
\phi^{\pr \tilde{z}\tilde{z}^\pr}(\mu=\mu^\pr=m={\cal M}), 
\,\Psi^{zz^\pr}({\cal M})\,\}\,,
\eea
where $z=-\tilde{z}=\pm 1$, $z^\prime=-\tilde{z}^\pr=\pm 1$.
Here $m,\mu$ and $\mu^\pr$ denote the bulk and brane masses
of scalar field, and ${\cal M}$ is the kink mass of 
Dirac fermion field.
We then find
\bea
\label{result2}
\l(\Delta_a\r)_{\rm SUSY} && = \l[ T_a({\cal V}^{++})
+ T_a({\cal V}^{--})\r]\ln (\Lambda \pi R)
\nonumber \\
&&-T_a({\cal H}^{++})\ln \l(\f{\Lambda(e^{{\cal M}_{++}\pi R}
-e^{-{\cal M}_{++}\pi R})}{2{\cal M}_{++}}\r)
\nonumber \\
&&-T_a({\cal H}^{+-})\ln \l(e^{-{\cal M}_{+-}\pi R}\r)
\nonumber \\
&&-T_a({\cal H}^{-+})\ln \l(e^{{\cal M}_{-+}\pi R}\r)
\nonumber \\
&&-T_a({\cal H}^{--})\ln \l(\f{\Lambda(e^{{\cal M}_{--}\pi R}
-e^{-{\cal M}_{--}\pi R})}{2{\cal M}_{--}}\r)
\eea
and the 4D beta function coefficients
\bea
\l(b_a\r)_{\rm SUSY}=-3T_a({\cal V}^{++})+T_a({\cal V}^{--})
+T_a({\cal H}^{++})+T_a({\cal H}^{--}).
\eea
In fact, one can obtain $(\Delta_a)_{\rm SUSY}$ 
using the 4D effective supergravity
method discussed in  \cite{Choi:2002wx,Choi:2002zi}.
We confirmed that the result from 4D effective supergravity agrees with 
the above result which was 
obtained from a direct calculation of KK threshold
correction.

\medskip

With the above result on $\Delta_a$, one can obtain
the low energy gauge couplings at $p\lesssim M_{KK}$
which are determined as 
(\ref{lowenergycoupling}) at one-loop approximation.
In most of 5D orbifold field theories, we have
\beq
\label{scalehierarchy}
\f{M_{KK}}{M_W}\,\gg \,\f{\Lambda}{M_{KK}}
\eeq
by many orders of magnitude, where $M_W$ is the weak scale
and $M_{KK}$ is the {\it lightest} KK mass.
Then the dominant part of higher order corrections
(beyond one-loop) to low energy couplings at $M_W$
come from the energy scales below $M_{KK}$, which
can be systematically computed within 4D effective theory.
To include those higher order corrections,
one can start with the matching condition
at $M_{KK}$:
\beq
\f{1}{g_a^2(M_{KK})}=\l(\f{1}{g_a^2}\r)_{\rm bare}
+\f{1}{8\pi^2}\l[
\Delta_a+b_a\ln\l(\f{\Lambda}{M_{KK}}\r)\r]\,,
\eeq
where $\Delta_a$ are given by (\ref{result1}),
and then subsequently perform 
two-loop renormalization group (RG) analysis
over the scales between $M_{KK}$ and $M_W$.
If there is a massive particle
with mass $M$ between $M_{KK}$ and $M_W$,
one needs to stop at $M$ to integrate out 
this massive particle, which would yield a new matching condition
at $M$.
For a given model, one can repeat this procedure to 
find the gauge couplings at $M_W$, and compare the results
with the experimentally measured values.

\medskip

In some case, there can be another large mass gap between
the lightest KK mass $M_{KK}$ and the {\it next} lightest
KK mass  ${M}^\pr_{KK}$.
For instance, as we have noticed in Sec. II, 
$\Psi^{+-}$ has two light KK modes with a Dirac mass
$2{\cal M}e^{-{\cal M}\pi R}$ when its kink mass 
${\cal M}\pi R\gg 1$.
Then the lightest KK mass is given by $M_{KK}=2{\cal M}e^{-{\cal M}\pi R}$,
while  the next lightest  KK mass $M^\pr_{KK}=1/R$
which corresponds to the mass of the first
KK mode of gauge fields.
If ${\cal M}\pi R$ is large enough, so that
\beq
\label{scalehierarchy2}
\f{{M}^\pr_{KK}}{M_{KK}}=
\f{e^{{\cal M}\pi R}}{2{\cal M}R}\,\gg\, \f{\Lambda}{{M}^\pr_{KK}}=
\Lambda R\,,
\eeq
the next important higher order corrections would come
from energy scales between $M_{KK}$ and $M^\pr_{KK}$. 
Those next important higher order corrections can be included
by performing the two-loop RG analysis starting from $M^\pr_{KK}$.
The corresponding matching condition
at $M_{KK}^\pr$ is given by
\beq
\f{1}{g^2_a(M^\pr_{KK})}=\l(\f{1}{g_a^2}\r)_{\rm bare}
+\f{1}{8\pi^2}\l[\Delta^\pr_a +b_a^\pr \ln\l(\f{\Lambda}{M^\pr_{KK}}
\r)\r]\,,
\eeq
where
\bea
b^\pr_a &=&b_a+\delta b_a\,,
\nonumber \\
\Delta^\pr_a &=& \Delta_a-\delta b_a \ln 
\l(\f{\Lambda}{M_{KK}}\r)
\eea
for $\Delta_a$ given by (\ref{result1}).
Here $b_a$ denote the one-loop beta function
coefficients for zero modes, while
$\delta b_a$ denote the 
coefficient for the lightest KK states.

\section{Application to orbifold GUT}

In the previous section, we have discussed one-loop 
gauge couplings in generic 5D orbifold field theory,
including the KK threshold correction $\Delta_a$ as
\beq
\label{relation}
\f{1}{g_a^2(p)}=\l(\f{1}{g_a^2}\r)_{\rm bare}
+\f{1}{8\pi^2}\l[\,\Delta_a+b_a\ln\f{\Lambda}{p}\,\r]
\quad\quad (a=1,2,3).
\eeq
The bare couplings here consist of several pieces
which are not calculable within orbifold field theory:
\beq
\l(\f{1}{g_a^2}\r)_{\rm bare}=
\f{\pi R}{g_{5a}^2}+\f{\gamma_a}{24\pi^3}\Lambda\pi R
+\kappa_a+\kappa^\pr_a.
\eeq
So although it is an well-defined 
relation between the bare parameters and measurable
quantities,  (\ref{relation}) does not give any useful
prediction {\it unless}  additional information on bare couplings
is provided.   
In orbifold GUT which is strongly coupled at $\Lambda$,
$g_{5a}^2$ and $\gamma_a$ are {\it universal}
as a consequence of unified gauge symmetry in bulk,
and both $\kappa_a$ and $\kappa^\pr_a$ are of the order of
$1/8\pi^2$ as the theory is strongly coupled at $\Lambda$ \cite{Chacko:1999hg,Nomura:2001tn}.
We then have
\beq
\l(\f{1}{g_a^2}\r)_{\rm bare}
=\f{1}{g^2_{\rm GUT}}+{\cal O}\l(\f{1}{8\pi^2}\r) \quad\quad (a=1,2,3)\,,
\eeq
With this information on bare couplings, 
one would be able to predict for instance
the value of QCD coupling constant at $M_Z$ in terms of
the measured values of the electroweak coupling constants at
$M_Z$ and the KK threshold corrections $\Delta_a$ computed in the
previous section.

\medskip

Let us discuss in more detail the effect of KK threshold 
on the {\it predicted}
value of the QCD coupling constant. 
To this end, it is convenient to consider 
\beq
\sum_a\f{\eta_a}{\alpha_a(p)}\,,
\eeq  
where $\alpha_a=g_a^2/4\pi$ and $\eta_a$ are the coefficients determined by
\beq
\eta_3=1, \quad \sum_a \eta_a=0, \quad
\sum_a \eta_a b_a =0.
\eeq
It is then straightforward to find
\beq
\f{1}{\alpha_3(M_Z)}\,=\,\l(\f{1}{\alpha_3(M_Z)}\r)_0
+\f{1}{2\pi}
\l(\eta_1\Delta_1+\eta_2\Delta_2+\Delta_3\r)
\eeq
where 
\beq
\l(\f{1}{\alpha_3(M_Z)}\r)_0=
-\l(\,\f{\eta_1}{\alpha_1(M_Z)}+\f{\eta_2}{\alpha_2(M_Z)}
+\delta_{\rm LE} \,\r)
\eeq
for the {\it experimentally measured} electroweak
coupling constants $\alpha_1(M_Z)=0.0169$ and $\alpha_2(M_Z)=0.0338$,
and
\beq
\delta_{\rm LE}=\sum_a\f{\eta_a}{\alpha_a(M_{KK})}
-\sum_a\f{\eta_a}{\alpha_a(M_Z)}
\eeq
can be determined by the RG analysis below $M_{KK}$
once the zero mode spectrums are known.
Note that when expanded in powers of $\Delta_a/8\pi^2$,
$\delta_{\rm LE}$ is {\it independent} of $\Delta_a$ at leading order.
If zero mode spectrums correspond to the standard
model (SM), we have
\beq
\eta_1= \f{115}{218}, \quad \eta_2= - \f{333}{218} 
\eeq
and 
\beq
\label{sm0}
\l(\f{1}{\alpha_3(M_Z)}\r)_0= 15.3 +{\cal O}\l(\f{1}{\pi}\r)
\eeq
for $M_{KK}=10^{13}-10^{15}$ GeV.
Here we consider a rather wide range of $M_{KK}$ to cover the
case that the lightest KK mass is suppressed by
a small localization factor as $M_{KK}=2{\cal M}e^{-{\cal M}\pi R}$.
It turns out that the numerical result is  insensitive
to $M_{KK}$, e.g. the variation of $1/\alpha_3$ is within ${\cal O}(1/\pi)$
even when $M_{KK}$ varies by several orders of magnitude.
In the above, we have used the approximation scheme to include two-loop RG
evolution below $M_{KK}$, and then
the uncertainty of ${\cal O}(1/\pi)$
is from the dependence of $\delta_{\rm LE}$ on
$M_{KK}$ as well as from the piece higher order in $\Delta_a/8\pi^2$.
For more interesting case that zero mode spectrums 
correspond to the minimal supersymmetric standard model (MSSM),
\beq
\eta_1= \f{5}{7}, \quad \eta_2= -\f{12}{7}
\eeq
and 
\beq
\label{mssm0}
\l(\f{1}{\alpha_3(M_Z)}\r)_0= 7.8 + {\cal
  O}\l(\f{1}{\pi}\r)\,.
\eeq
Here we have assumed the superparticle masses $M_{\rm SUSY}$
at $0.3 \sim 1$ TeV,
and then the uncertainty of ${\cal O}(1/\pi)$ is mainly from
the variation of superparticle threshold effects at $M_{\rm SUSY}$.

\medskip
To see the importance of
KK threshold corrections more explicitly,
let us consider a class of 
5D $SU(5)$ orbifold GUTs whose effective 4D theory 
is given by the MSSM.
To break $SU(5)$ by orbifolding,
$Z_2\times Z_2^\pr$ is embedded into $SU(5)$ as 
\bea
{Z_2} = {\rm
    diag}(+1,+1,+1,+1,+1)\,, \nonumber \\
Z^\pr_2 = {\rm
    diag}(+1,+1,+1,-1,-1)\,,  
\eea
leading to the following orbifold boundary conditions
of the 5D vector multiplet $V$ containing the $SU(5)$ gauge
fields:
\beq
V=(8,1)_0^{(++)}+(1,3)_0^{(++)}+(1,1)_0^{(++)}
+(3,2)_{-5/6}^{(+-)}+(\bar{3},2)_{5/6}^{(+-)}\,,
\nonumber 
\eeq
where the $SU(5)$ adjoint representation is decomposed
into the representations of the SM gauge group.
Obviously then the bulk $SU(5)$ is broken down to $SU(3)\times
SU(2)\times U(1)$ at $y=\pi R$, while it is unbroken
at $y=0$. 

\medskip

The model contains matter hypermultiplets 
$F_p(\bar{5})$, $F^\pr_p(\bar{5})$, $T_p(10)$ and 
$T^\pr_p(10)$ ($p=1,2,3$) 
with kink masses
${\cal M}_{F_p}$, ${\cal M}_{F^\pr_p}$, 
${\cal M}_{T_p}$ and ${\cal M}_{T^\pr_p}$, 
and also the Higgs hypermultiplets $H(5)$ and $H^\pr(\bar{5})$
with kink masses 
${\cal M}_H$ and
${\cal M}_{H^\pr}$,
where the numbers in bracket mean the $SU(5)$ representation.
We assign the $Z_2\times Z_2^\pr$ parities of these hypermultiplets
as
\bea
\label{bc2}
&&Z_2(F_p)=Z_2(F^\pr_p)=Z_2(T_p)=Z_2(T^\pr_p)=Z_2(H)=Z_2(H^\pr)=1
\nonumber \\
&& Z_2^\pr(F_p)=-Z_2^\pr(F^\pr_p)=
Z^\pr_2(T_p)=-Z^\pr_2(T^\pr_p)=Z^\pr_2(H)=Z_2^\pr(H^\pr)
=-1\,,
\eea
and then the orbifold
boundary conditions of matter and Higgs hypermultiplets
are given by
\bea
&& F=(\bar{3},1)_{1/3}^{(+-)}+(1,2)_{-1/2}^{(++)}\,,
\nonumber \\
&& F^\pr=(\bar{3},1)_{1/3}^{(++)}+(1,2)_{-1/2}^{(+-)}\,,
\nonumber \\
&& T=(3,2)_{1/6}^{(++)}+(\bar{3},1)_{-2/3}^{(+-)}
+(1,1)_{1}^{(+-)}\,,\nonumber \\
&& T^\pr=(3,2)_{1/6}^{(+-)}+
(\bar{3},1)_{-2/3}^{(++)}+
(1,1)_{1}^{(++)}\,,\nonumber \\
&& H=(3,1)_{-1/3}^{(+-)}+(1,2)_{1/2}^{(++)}\,,
\nonumber \\
&& H^\pr=(\bar{3},1)_{1/3}^{(+-)}+
(1,2)_{-1/2}^{(++)}\,.
\eea
Then using (\ref{mssm0}) and (\ref{result2}), we find
\bea
\f{1}{\alpha_3 (M_Z)} 
 = 7.8+\f{1}{2\pi}\l[\,\Delta_{\rm gauge}
+\Delta_{\rm higgs}+\Delta_{\rm matter}\r]+{\cal O}\l(
\f{1}{\pi}\r)
\eea
where  
\bea
\f{1}{2\pi}\Delta_{\rm gauge}=\f{3}{7\pi}\,\ln (\pi R\Lambda)
\eea
corresponds to the KK threshold correction from
the 5D vector multiplet,
\bea
\label{higgs}
\Delta_{\rm higgs}=\f{9}{14}\, \l[\,\ln \l(
\f{\sinh \,\pi R{\cal M}_{H}}{\pi R {\cal M}_{H}}\r)
+\pi R {\cal M}_H 
+ \ln 
\l(\f{\sinh \,\pi R {\cal M}_{H^\pr}}{\pi R {\cal M}_{H^\pr}}\r)+ 
\pi R{\cal M}_{H^\pr}  \right]
\eea
is the correction from Higgs hypermultiplets, and
\bea
\label{matter}
\Delta_{\rm matter}&=&
\f{9}{14} \sum_p\l[\,\ln \l(\f{\sinh \,\pi R{\cal M}_{F_p}}
{\pi R{\cal M}_{F_p}}\r)
+\pi R{\cal M}_{F_p}
-\ln \l(\f{\sinh \,\pi R{\cal M}_{F^\pr_p}}{\pi R{\cal M}_{F^\pr_p}}\r)
-\pi R {\cal M}_{F^\pr_p}\,\r]
\nonumber \\
&+&\f{3}{2} \sum_p \l[\,
\ln\l(\f{\sinh \, \pi R{\cal M}_{T_p}}{\pi R{\cal M}_{T_p}}\r) 
+\pi R{\cal M}_{T_p} -
 \ln\l(\f{\sinh \, \pi R {\cal M}_{T^\pr_p}}{\pi R{\cal M}_{T^\pr_p}}\r) 
-\pi R{\cal M}_{T^\pr_p}\,\r]
\label{a3diff}
\eea 
is the correction from matter hypermultiplets.

\medskip

For $\Lambda\pi R\approx 10^2$, 
we have
\beq
\f{1}{2\pi}\Delta_{\rm gauge}\approx 0.8\,.
\eeq
Then the predicted
value of $1/\alpha_3$ would be {\it very close} to the experimental value:
\beq
\l(\f{1}{\alpha_3(M_Z)}\r)_{\rm exp}=8.55\pm 0.15\,,
\eeq
{if} $\Delta_{\rm higgs}+\Delta_{\rm matter}$ is negligible.
In other words,
the KK threshold correction from Higgs and matter
hypermultiplets should be negligible, i.e.
\beq
\label{condition}
\f{1}{2\pi}\l(\,\Delta_{\rm higgs}+\Delta_{\rm matter}\,\r)
\,\lesssim\, {\cal O}\l(\f{1}{\pi}\r)\,,
\eeq
in order for 
the 5D orbifold GUT under consideration to be consistent
with observation.

\medskip 

Obviously, for the class of models under consideration,
a hypermultiplet with ${\cal M}\pi R\gg 1$
gives a large threshold correction 
$\Delta (1/\alpha_3)={\cal O}({\cal M}\pi R/2\pi)$
which can make the model inconsistent with the
observation.
For instance, in the model of \cite{Hebecker:2002re}, 
the Higgs zero modes are 
quasi-localized at $y=0$ by having
${\cal M}_{H}\pi R= 11.5$ and ${\cal M}_{H^\pr}\pi R = 6.9$,
the 1st and 3rd generation matters are brane
fields at $y=\pi R$ and $y=0$, respectively,
and the 2nd generation matters come from
hypermultiplets with vanishing kink masses.
This model then gives
$1/\alpha_3 (M_Z) = 10.9 $
which is too large to be consistent with the experimental value.

\medskip

A simple way to avoid a too large $\Delta_{\rm higgs}$
is to assume that both of the Higgs hypermultiplets
have  ${\cal M}\pi R\ll -1$, for instance
${\cal M}_H\pi R\approx {\cal M}_{H^\pr} \pi R\approx -10$. 
In this case, the Higgs zero modes are localized at $y=\pi R$, and
\beq
\frac{1}{2\pi}\Delta_{\rm higgs}\approx-\f{9}{14\pi}\,\ln\l(
\sqrt{{\cal M}_{H}{\cal M}_{H^\pr}}\,\pi R\r)
\approx -0.45,
\eeq
which is small enough {\it not} to spoil the successful
gauge unification.
However for matter hypermultiplets,
to generate the hierarchical 4D Yukawa couplings through
dynamical quasi-localization, one needs to localize heavy 
and light generations at different locations.
This means that some kink masses should be
positive, while some others are negative.
One then needs a {\it nontrivial cancellation} between the corrections
from different matter hypermultiplets
in order for $\Delta_{\rm matter}\approx 0$.
In this regard, an interesting possibility is that
\beq
\label{su2}
{\cal M}_{F_p}={\cal M}_{F^\pr_p}\,,
\quad 
{\cal M}_{T_p}={\cal M}_{T^\pr_p}\,,
\eeq
which obviously lead to
\beq
\f{1}{2\pi}\Delta_{\rm matter}= 0\,.
\eeq
The relation (\ref{su2}) between hypermultiplet masses 
can be considered as a consequence of
global $SU(2)_H$ symmetry under which $(F_p,F^\pr_p)$ and
$(T_p,T^\pr_p)$ transform as a doublet.
This $SU(2)_H$ symmetry is 
broken down to $U(1)_{3B+L}$ at $y=\pi R$
by the orbifolding boundary conditions
(\ref{bc2}).
It is then straightforward to introduce a dynamics
at $y=\pi R$ which breaks $U(1)_{3B+L}$ 
spontaneously down to the matter-parity
$(-1)^{3B+L}$. This is in fact necessary
to generate nonzero Majorana masses of neutrinos.
Then the brane interactions at $y=\pi R$
are constrained only by the SM gauge group, 4D $N=1$ supersymmetry
and the $R$-parity $(-1)^{3B+L+2s}$ ($s=$ spin).

\medskip

\begin{table}
\caption{A set of
  hypermultiplet masses which give
  realistic fermion masses and CKM mixing while
satisfying (\ref{condition}) for successful gauge unification.
Here $n=0,1,2$.}
\begin{tabular}{c|c|c|c}
\hline
\hline
${\cal M}_{T_p} / M_0$ &${\cal M}_{T'_p}/ M_0$ & ${\cal M}_{F_p} /M_0$ &
${\cal M}_{F'_p} / M_0$ \\
\hline
\hline
~ (4,2,0) & (4,2,0) & (n,n,n) & (n,n,n)  \\
\hline
~ (5,1,0) & (3,3,0) & (2,0,1) & (1,2,0) \\
\hline
~ (3,3,0) & (5,1,0) & (0,2,1) & (2,0,1) \\
\hline
~ (6,6,0) & (2,4,0) & (4,0,2) & (0,4,2) \\
\hline
~ (5,1,0) & (3,3,0) & (3,1,2) & (1,3,2) \\
\hline
~ (3,3,0) & (3,3,0) & (1,3,2) & (3,1,2) \\
\hline
~ (2,4,0) & (6,0,0) & (0,4,2) & (4,0,2) \\
\hline
\hline
\end{tabular}
\end{table}

The KK threshold corrections 
(\ref{higgs}) and (\ref{matter}) to $1/\alpha_3$
from Higgs and matter hypermultiplets have a correlation
with the 4D Yukawa couplings given by
(\ref{canonicalyukawa}).
So the condition (\ref{condition})
for successful gauge unification provides
some restriction on the possible forms of 4D Yukawa couplings.
However still it is not so difficult to construct
models to produce the correct form of Yukawa couplings
through quasi-localization, while satisfying
(\ref{condition}).
To see this, consider a class of models with
${\cal M}_H\pi R\ll -1$ and
${\cal M}_{H^\pr}\pi R\ll -1$, in which the Higgs zero modes are
quasi-localized at $y=\pi R$. 
The quark and lepton Yukawa couplings
are assumed to arise
from the following brane interactions at $y=\pi R$:
\beq
\int d^5x \int d^2\theta \,\delta (y-\pi R)\f{1}{\Lambda^{3/2}}
\l[\, \lambda^U_{pq}H Q_pU^c_q+
\lambda^D_{pq}H' Q_pD^c_q+\lambda^E_{pq}H' L_pE^c_q\,\r]\,,
\eeq
where the Higgs doublets $H_1$ and $H_2$ come from
$H$ and $H^\pr$, respectively, 
the lepton doublets $L_p$ are from $F_p$,
the lepton singlets $E^c_p$ are from $T^\pr_p$,
the quark doublets $Q_p$ are from $T_p$,
the quark singlets $U^c_p$ and $D^c_p$ are from
$T^\pr_p$ and $F^\pr_p$, respectively.
Then according to the discussion of Sec. II, 
the physical Yukawa couplings
of quarks and leptons are given by
\bea
&&y^U_{pq}=\sqrt{Z({\cal M}_H)Z({\cal M}_{T^\pr_p})
Z({\cal M}_{T_q})}\,\lambda^U_{pq}\,,\nonumber \\
&&y^D_{pq}=\sqrt{Z({\cal M}_{H'})Z({\cal M}_{T^\pr_p})
Z({\cal M}_{F^\pr_q})}\,\lambda^D_{pq}\,,\nonumber \\
&&y^E_{pq}=\sqrt{Z({\cal M}_{H'})Z({\cal M}_{F_p})
Z({\cal M}_{T_q})}\,\lambda^E_{pq}\,,
\eea
where
\beq
Z({\cal M})=\f{2{\cal M}}{\Lambda}\f{1}{e^{2{\cal M}\pi R}-1}.
\eeq
If we further assume that all hypermultiplet masses are quantized
in an appropriate unit, then the hypermultiplet masses 
tabulated in Table I give the correct quark and lepton masses as well as 
the correct CKM mixing angles, while satisfying
(\ref{condition}) for successful gauge unification.
Here the unit mass ${\cal M}_0$ is defined to given by
$e^{-{\cal M}_0\pi R}=$ Cabbibo angle $\approx 0.2$.

\section{conclusion}

In this paper, we have examined the KK threshold corrections
to low energy gauge couplings $g_a^2$ from bulk matter fields
whose zero modes are dynamically quasi-localized to generate
hierarchical 4D Yukawa couplings. 
We derived the explicit form of threshold corrections
in generic 5D orbifold field theory on $S^1/Z_2\times
Z_2^\pr$, and found that the corrections to
$1/g_a^2$ are of the order of
$\ln (y)/8\pi^2$ where $y$ denotes 4D Yukawa couplings generated by
quasi-localization.
So generically quasi-localization significantly affects
gauge coupling unification. 
We then applied the results to 5D orbifold GUT, and
discussed the conditions for a 5D GUT to 
generate hierarchical Yukawa couplings without spoiling
successful gauge coupling unification.
Some examples of such 5D GUTs are presented in Table I.

\begin{acknowledgments}
This work is supported by KRF PBRG 2002-070-C00022.
\end{acknowledgments}


\begin{thebibliography}{99}
\bibitem{Arkani-Hamed:1999dc}
N.~Arkani-Hamed and M.~Schmaltz,
Phys.\ Rev.\ D {\bf 61}, 033005 (2000)
[arXiv:hep-ph/9903417].


\bibitem{Mirabelli:1999ks}
E.~A.~Mirabelli and M.~Schmaltz,
Phys.\ Rev.\ D {\bf 61}, 113011 (2000)
[arXiv:hep-ph/9912265].

\bibitem{Dvali:2000ha}
G.~R.~Dvali and M.~A.~Shifman,
Phys.\ Lett.\ B {\bf 475}, 295 (2000)
[arXiv:hep-ph/0001072].

\bibitem{Kaplan:2000av}
D.~E.~Kaplan and T.~M.~Tait,
JHEP {\bf 0006}, 020 (2000)
[arXiv:hep-ph/0004200].

\bibitem{Arkani-Hamed:2001tb}
N.~Arkani-Hamed, T.~Gregoire and J.~Wacker,
JHEP {\bf 0203}, 055 (2002)
[arXiv:hep-th/0101233].


\bibitem{Kaplan:2001ga}
D.~E.~Kaplan and T.~M.~Tait,
JHEP {\bf 0111}, 051 (2001)
[arXiv:hep-ph/0110126].

\bibitem{Kakizaki:2001ue}
M.~Kakizaki and M.~Yamaguchi,
arXiv:hep-ph/0110266.


\bibitem{Haba:2002uw}
N.~Haba and N.~Maru,
Phys.\ Rev.\ D {\bf 66}, 055005 (2002)
[arXiv:hep-ph/0204069].

\bibitem{Hebecker:2002re}
A.~Hebecker and J.~March-Russell,
Phys.\ Lett.\ B {\bf 541}, 338 (2002)
[arXiv:hep-ph/0205143].

\bibitem{Grossman:2002pb}
Y.~Grossman and G.~Perez,
Phys.\ Rev.\ D {\bf 67}, 015011 (2003)
[arXiv:hep-ph/0210053].

\bibitem{Chang:2002ww}
W.~F.~Chang and J.~N.~Ng,
JHEP {\bf 0212}, 077 (2002)
[arXiv:hep-ph/0210414].

\bibitem{Kitano:2003cn}
R.~Kitano and T.~j.~Li,
Phys.\ Rev.\ D {\bf 67}, 116004 (2003)
[arXiv:hep-ph/0302073].

\bibitem{Choi:2003di}
K.~Choi, D.~Y.~Kim, I.~W.~Kim and T.~Kobayashi,
arXiv:hep-ph/0305024.

\bibitem{Biggio:2003kp}
C.~Biggio, F.~Feruglio, I.~Masina and M.~Perez-Victoria,
arXiv:hep-ph/0305129.

\bibitem{Kawamura:2000ev}
Y.~Kawamura,
Prog.\ Theor.\ Phys.\  {\bf 105}, 999 (2001)
[arXiv:hep-ph/0012125].

\bibitem{Altarelli:2001qj}
G.~Altarelli and F.~Feruglio,
Phys.\ Lett.\ B {\bf 511}, 257 (2001)
[arXiv:hep-ph/0102301].

\bibitem{Hall:2001pg}
L.~J.~Hall and Y.~Nomura,
Phys.\ Rev.\ D {\bf 64}, 055003 (2001)
[arXiv:hep-ph/0103125];
Phys.\ Rev.\ D {\bf 65}, 125012 (2002)
[arXiv:hep-ph/0111068];
Phys.\ Rev.\ D {\bf 66}, 075004 (2002)
[arXiv:hep-ph/0205067].



\bibitem{Hebecker:2001wq}
A.~Hebecker and J.~March-Russell,
Nucl.\ Phys.\ B {\bf 613}, 3 (2001)
[arXiv:hep-ph/0106166].

\bibitem{Hebecker:2001jb}
A.~Hebecker and J.~March-Russell,
Nucl.\ Phys.\ B {\bf 625}, 128 (2002)
[arXiv:hep-ph/0107039].

\bibitem{Hall:2001tn}
L.~J.~Hall, H.~Murayama and Y.~Nomura,
Nucl.\ Phys.\ B {\bf 645}, 85 (2002)
[arXiv:hep-th/0107245].

\bibitem{Asaka:2001eh}
T.~Asaka, W.~Buchmuller and L.~Covi,
Phys.\ Lett.\ B {\bf 523}, 199 (2001)
[arXiv:hep-ph/0108021].

\bibitem{Hall:2001xr}
L.~J.~Hall, Y.~Nomura, T.~Okui and D.~R.~Smith,
Phys.\ Rev.\ D {\bf 65}, 035008 (2002)
[arXiv:hep-ph/0108071].

\bibitem{Dermisek:2001hp}
R.~Dermisek and A.~Mafi,
Phys.\ Rev.\ D {\bf 65}, 055002 (2002)
[arXiv:hep-ph/0108139].

\bibitem{Kim:2002ab}
H.~D.~Kim, J.~E.~Kim and H.~M.~Lee,
JHEP {\bf 0206}, 048 (2002)
[arXiv:hep-th/0204132].

\bibitem{Kim:2002im}
H.~D.~Kim and S.~Raby,
JHEP {\bf 0301}, 056 (2003)
[arXiv:hep-ph/0212348].
H.~D.~Kim and S.~Raby,
JHEP {\bf 0307}, 014 (2003)
[arXiv:hep-ph/0304104].





\bibitem{Weinberg:1980wa}
S.~Weinberg,
Phys.\ Lett.\ B {\bf 91}, 51 (1980);
L.~J.~Hall,
Nucl.\ Phys.\ B {\bf 178}, 75 (1981).

\bibitem{Choi:1987iu}
K.~Choi,
Phys.\ Rev.\ D {\bf 37}, 1564 (1988);
V.~S.~Kaplunovsky,
Nucl.\ Phys.\ B {\bf 307}, 145 (1988)
[Erratum-ibid.\ B {\bf 382}, 436 (1992)]
[arXiv:hep-th/9205068].




\bibitem{Gherghetta:2000qt}
T.~Gherghetta and A.~Pomarol,
Nucl.\ Phys.\ B {\bf 586}, 141 (2000)
[arXiv:hep-ph/0003129].

\bibitem{Chacko:1999hg}
Z.~Chacko, M.~A.~Luty and E.~Ponton,
JHEP {\bf 0007}, 036 (2000)
[arXiv:hep-ph/9909248].

\bibitem{Nomura:2001tn}
Y.~Nomura,
Phys.\ Rev.\ D {\bf 65}, 085036 (2002)
[arXiv:hep-ph/0108170].

\bibitem{Froggatt:1978nt}
C.~D.~Froggatt and H.~B.~Nielsen,
Nucl.\ Phys.\ B {\bf 147}, 277 (1979).

\bibitem{Ceresole:2000jd}
A.~Ceresole and G.~Dall'Agata,
Nucl.\ Phys.\ B {\bf 585}, 143 (2000)
[arXiv:hep-th/0004111].


\bibitem{Choi:2002wx}
K.~Choi, H.~D.~Kim and I.~W.~Kim,
JHEP {\bf 0211}, 033 (2002)
[arXiv:hep-ph/0202257].

\bibitem{Contino:2001si}
R.~Contino, L.~Pilo, R.~Rattazzi and E.~Trincherini,
Nucl.\ Phys.\ B {\bf 622}, 227 (2002)
[arXiv:hep-ph/0108102].

\bibitem{Choi:2002zi}
K.~Choi, H.~D.~Kim and I.~W.~Kim,
JHEP {\bf 0303}, 034 (2003)
[arXiv:hep-ph/0207013].

\bibitem{Choi:2002ps}
K.~Choi and I.~W.~Kim,
Phys.\ Rev.\ D {\bf 67}, 045005 (2003)
[arXiv:hep-th/0208071].

\bibitem{Arkani-Hamed:2001is}
N.~Arkani-Hamed, A.~G.~Cohen and H.~Georgi,
Phys.\ Lett.\ B {\bf 516}, 395 (2001)
[arXiv:hep-th/0103135];
C.~A.~Scrucca, M.~Serone, L.~Silvestrini and F.~Zwirner,
Phys.\ Lett.\ B {\bf 525}, 169 (2002)
[arXiv:hep-th/0110073].


\bibitem{GrootNibbelink:2001bx}
S.~Groot Nibbelink,
Nucl.\ Phys.\ B {\bf 619}, 373 (2001)
[arXiv:hep-th/0108185];
R.~Contino and A.~Gambassi,
J.\ Math.\ Phys.\  {\bf 44}, 570 (2003)
[arXiv:hep-th/0112161].






\end{thebibliography}
\end{document}